\documentclass[preprint,pre,aps]{revtex4}

\usepackage{graphicx}
\usepackage{amsmath}
\usepackage{amsfonts}
\usepackage{amssymb}
\usepackage{epsf}
\usepackage{bm}

\begin{document}

\title{Speeding chemical reactions by focusing}

\author{ A. M. Lacasta$^1$, L. Ram\'irez--Piscina$^1$, J. M. Sancho$^2$ and  K. Lindenberg$^3$}
 \affiliation{
$^1$Departament de F\'{\i}sica Aplicada, Universitat Polit\`ecnica de Catalunya,\\ Avinguda Doctor Mara\~n\'on 44, E-08028 Barcelona, Spain\\
$^2$Departament d'Estructura i Constituents de la Mat\`eria,\\
Facultat de F\'{\i}sica, Universitat de Barcelona,\\
Carrer Mart\'i i Franqu\'es 1, E-08028 Barcelona, Spain\\
$^3$Department of Chemistry and Biochemistry and BioCircuits Institute,\\
University of California San Diego,\\
9500 Gilman Drive, La Jolla, CA 92093-0340, USA\\
}
\date{\today}

\begin{abstract}
 We present numerical results for a chemical reaction of colloidal particles  which  are transported 
 by a laminar fluid  and are focused by  periodic 
obstacles in such a way that the two components are well mixed and consequently the chemical reaction is speeded up.
 The roles of the various system parameters (diffusion coefficients, reaction rate, obstacles sizes) are studied. 
We show that focusing speeds up the reaction from the diffusion limited rate $\sim t^{-1/2}$ to very close to the perfect mixing rate, $\sim t^{-1}$.
\end{abstract}

\maketitle

\section{Introduction}

The motion of colloidal particles on modulated surfaces has attracted a great deal of attention in the past decade \cite{macdonald03,grier03,korda02,huang04,Morton,Xuan}. The interest in this subject has mainly been directed at sorting phenomena, and a considerable portion of the work has been experimental.  The work has focused on mixtures of particles which are sorted into separate streams of different species when the mixture is transported under laminar conditions along a structured or modulated medium with periodic obstacles or traps. In this scenario it is possible to control the  transport of materials  such as DNA fragments or functionalized biological colloidal particles.  The modulated surfaces are specifically designed to present periodic arrays of traps~\cite{grier03,korda02} or microfabricated obstacles~\cite{huang04} among other configurations. This technique can be applied not only to solid spherical particles but also to other objects such as cells, proteins, DNA, and  droplets in inmimiscible fluids~\cite{Champagne}.   Theoretical studies complemented with stochastic simulations have received considerable attention \cite{katja2,ana,James}. Other sorting methods based on inertia and hydrodynamics have also been explored~\cite{DiCarlo}. The sorting phenomenon consists of a lateral or orthogonal  displacement of the particles with respect to the driving force or velocity direction of the fluid mixture.  The deviation of the particles in a mixture from the direction of flow of the mixture depends on some property or group of properties of the particles such as, for instance, size, mass, or charge, causing the particles with different values of these properties to flow in different directions. 

The same principle can be used to achieve the converse effect, namely, to focus particles coming from different directions and mix them if the modulated structure of obstacles or traps is prepared accordingly~\cite{Morton}. Focusing of  particles is useful in a number of different scenarios such as counting, detecting, and mixing~\cite{Xuan}. This property has special relevance in the laminar regime, where slow molecular diffusion makes it difficult to concentrate and mix particles.

Our particular focus in this work lies in using this methodology to mix  reactants in order to speed their reaction. It is our objective to show that using an appropriate modulated or structured surface of obstacles one can concentrate two reactants in a very small domain, thus favoring their chemical reaction. Our main result is that reactants that arise from 
non-homogeneous distributions can be efficiently mixed and as a result are able to reach the classical law of mass action reaction regime characterized by the reactant concentration decay law $ \sim t^{-1}$ much sooner than they would in the absence of a mixing mechanism. We present results on the efficiency of some obstacle geometries toward this purpose. We discuss the roles of the different control parameters and of the densities, the diffusion coefficients of the different species, the reaction rates, and the particle and obstacle sizes.

Our presentation proceeds as follows.  First in Sec.~\ref{sec2} we begin with the description of the continuous dynamical scenario and present the associated dynamical equations which contain diffusion, advection and reaction terms. In Sec.~\ref{sec3}  we present  numerical results from the simulation of the equations and thereby analyze  the roles of the different parameters. Finally we close with some conclusions and perspectives in Sec.~\ref{sec4}.

\section{Dynamical model}
\label{sec2}

\begin{figure}
\begin{center}
\includegraphics[width = 12cm]{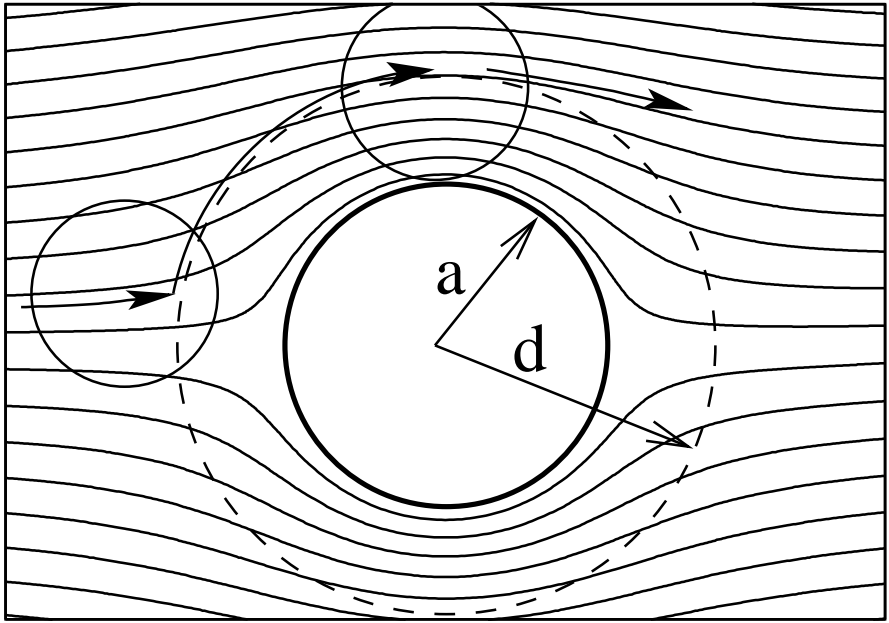}
\end{center}
\caption{Stream lines of the velocity field (see Appendix), one obstacle (central circle) and the motion (arrows)  of a particle (small circle) close to the obstacle. The dashed circle indicates the area of the obstacle's influence due to the finite size of the particle. The radius of the obstacle $a$ and the potential $d$ are indicated.}
\label{lines} 
\end{figure}

The theoretical scenario is the advective reaction--diffusion model of chemical kinetics corresponding to the simple irreversible reaction  $A+B \rightarrow 0$. The dynamical equations for the two reactants are
\begin{subequations}
\label{eq:1}
\begin{align}
\label{eq:1a}
 \frac{\partial}{\partial t} c_a(x,y;t) & = -\nabla {\mathbf J}_a(x,y;t) - k c_a c_b, \\ 
\frac{\partial}{\partial t} c_b(x,y;t) & = -\nabla {\mathbf J}_b(x,y;t) - k c_a c_b,
\label{eq:1b} 
\end{align}
\end{subequations}
where $c_a$ and $c_b$ are the time dependent  local concentrations of the reactants $A$ and $B$, $k$ is
 the reaction rate constant, and ${\bf J}_a$, ${\bf J}_b$ are the  fluxes of the reactants. The latter are given by
\begin{subequations}
\begin{align}
{\mathbf J}_a(x,y;t) &= c_a(x,y;t) {\mathbf v}(x,y;t)
  -D_a \nabla c_a 
  -U_0  c_a\nabla U, \\ 
{\mathbf J}_b(x,y;t) &= c_b(x,y;t) {\mathbf v}(x,y;t)
  -D_b \nabla c_b 
  -U_0  c_b\nabla U.
\end{align}
\end{subequations}
Here $D_a$ and $D_b$ are the diffusion coeficients, $U_0 U (x/\lambda, y/\lambda)$ is the modulated potential
 interaction due to obstacles, and ${\mathbf v}( x/\lambda, y/\lambda)$ is the local velocity
 responsible for the advective flux, which we assume to be a Hele-Shaw flow (that is, a flow between two very
 close parallel plates). We have explicitly extracted the amplitude $U_0$ of the potential so that $U(x/\lambda,y/\lambda)$ is
 the potential of unit amplitude. This potential is modeled by placing a circular tower at each obstacle, which changes from unit value at the center of the disk defining the base of the obstacle to a zero value outside the range of the interaction. Specifically, it is modeled by the expression
\begin{equation}
U({\bf r}) = 
\sum_{k=1}^N \frac{1}{2}\left(1 - \tanh{\frac{\left|{\bf r}-{\bf R}_k \right| - d}{\delta}} \right),
\end{equation}
where ${\bf R}_k$, $k=1 \dots N$, are the positions of the centers of the $N$ obstacles of base radius $a$, $d>a$ is the radius of the interaction, and $\delta$ is the (small) scale that characterizes a substantial change in the value of the potential. 
To take into account the finite size of the advected particles in a model with a continuous concentration field we have introduced an interaction potential with a radius $d$ larger than the obstacle radius $a$. The distance $d-a$ then represents the particle radius (see Fig.~\ref{lines}). The potential range $d$ corresponds to the minimum distance between the centers of the colloidal particle and the obstacle, and hence is the sum of their radii.
The flow field is obtained by solving the Laplace equation with boundary 
conditions that reflect the presence of the $N$ circular obstacles of radius $a$, 
but neglecting any effects that the advected particles might have on the flow. 
Even this is a rather arduous task that we have moved to the Appendix.

Equations~(\ref{eq:1}) can be simplified with a change to the new variables $\tau$, $x'$, and $y'$,
\begin{equation}
t = t_0 \tau, \qquad x= \lambda x', \qquad y= \lambda y',
\end{equation}
where $\lambda$ and $t_0$ are characteristic length and time scales. This transforms Eqs.~(\ref{eq:1}) to
\begin{subequations}
\begin{align}
\frac{\partial}{\partial \tau} c_a(x',y',\tau)& = {\hat D}_a \nabla^2 c_a +{\hat U}_0 \nabla (c_a\nabla U) + \nabla ({\mathbf  v} c_a) - {\hat k}  c_a c_b, \\ 
\frac{\partial}{\partial \tau} c_b(x',y',\tau)& = {\hat D}_b \nabla^2 c_b +{\hat U}_0 \nabla (c_b\nabla U) +  \nabla ({\mathbf  v} c_b) - {\hat k} c_a c_b. 
\end{align}
\label{eq:1-b}
\end{subequations}
The dimensionless parameters are given by
\begin{equation}
{\hat D_i} = \frac{D_i t_0}{\lambda^2}, \qquad {\hat U}_0= \frac{U_0 t_0}{\lambda^2}, \qquad  
{\hat k} = \frac{k t_0}{\lambda^3},
\end{equation}
and the concentrations and derivative operators are also dimensionless.
 We fix the parameters ${\hat D}_i=0.01$ and ${\hat U}_0=0.01$ throughout the paper.
 From now on for simplicity of notation we drop the primes on $x'$ and $y'$ and alert the 
reader not to confuse $x$ and $y$ as used henceforth with the original variables.

The equations are simulated on a two-dimensional lattice of 
$N_x=1000, N_y=300$ cell centers and cell dimensions (dimensionless quantities) $\Delta x= \Delta y= 0.05$, which corresponds to system dimensions $l_x=50$ and  $l_y=15$. Our integration
time step is $\Delta \tau=5 \times10^{-4}$. The dynamical evolution is reported  every $5$ time units up to a final time $\tau =60$. 

\begin{figure}
\begin{center}
\includegraphics[width = 14cm]{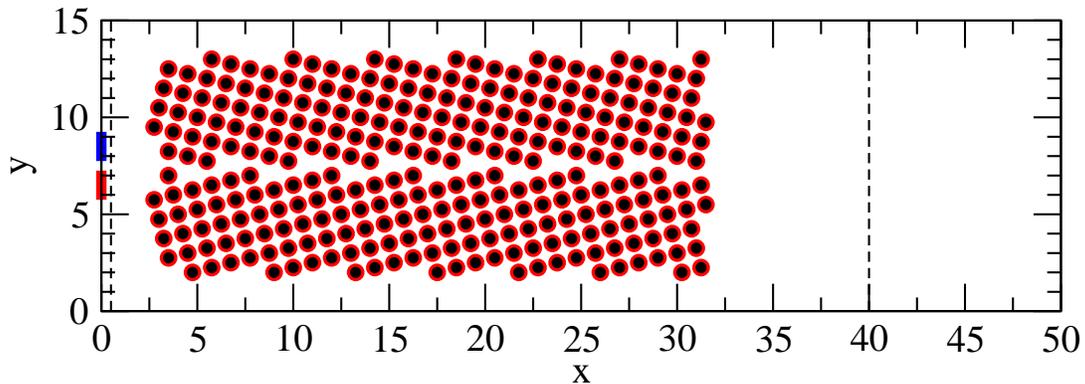}
\end{center}
\caption {
Periodic  pattern (PP) with  tilted obstacles. The black circles are the obstacles and the red area corresponds to the range of the interaction potential. The dashed lines show the positions where incoming and
final flows respectively are numerically measured. The blue and red bars on the $y$ axis denote the inlets through which the two reacting especies are introduced into the system (see Fig.\ \ref{2evol_I}).
}
\label{1evol_I}
\end{figure}

Figure \ref{lines} shows the role of one obstacle, the flow lines and the finite size of a colloidal particle. When the particle following a flux line is close to the obstacle its trajectory changes to a different  flow line pointing away from the obstacle. The result is a lateral deviation of the particle from the initial flow line as it circumnavigates the obstacle.
In the presence of several obstacles, and depending on their specific spatial distribution, these deviations can favor focusing and hence accelerate the reaction. To check this hypothesis we have employed two obstacle patterns,  one with 293 obstacles in a tilted periodic pattern (PP), as shown in Fig. \ref{1evol_I}, and a second one with the same number of obstacles but distributed randomly over the same area (random pattern, RP). We have also performed some simulations with no pattern (NP). We have mainly used obstacles of radius $a=0.15$ and an interaction potential radius $d=0.25$, which corresponds to a radius $d-a=0.10$ of reacting particles. We have also used the values $a=0.10$, $0.05$ and $d=0.20$ in some cases in order to study the dependence of the results on these parameters. We have also performed some simulations in which the flow is not affected by the obstacles and thus the particles move at a constant velocity, {\it i.e.}, ${\bf v} =$ constant ($a=0)$, see below. This is done because in previous calculations concerning flows of particle mixtures over surfaces with obstacles we did not take the effect of the obstacles on the flow field into account \cite{katja2,ana,James}.  Here we have the opportunity to assess the importance of doing so (see below).

\begin{figure}
\begin{center}
\includegraphics[width = 5cm,angle=90]{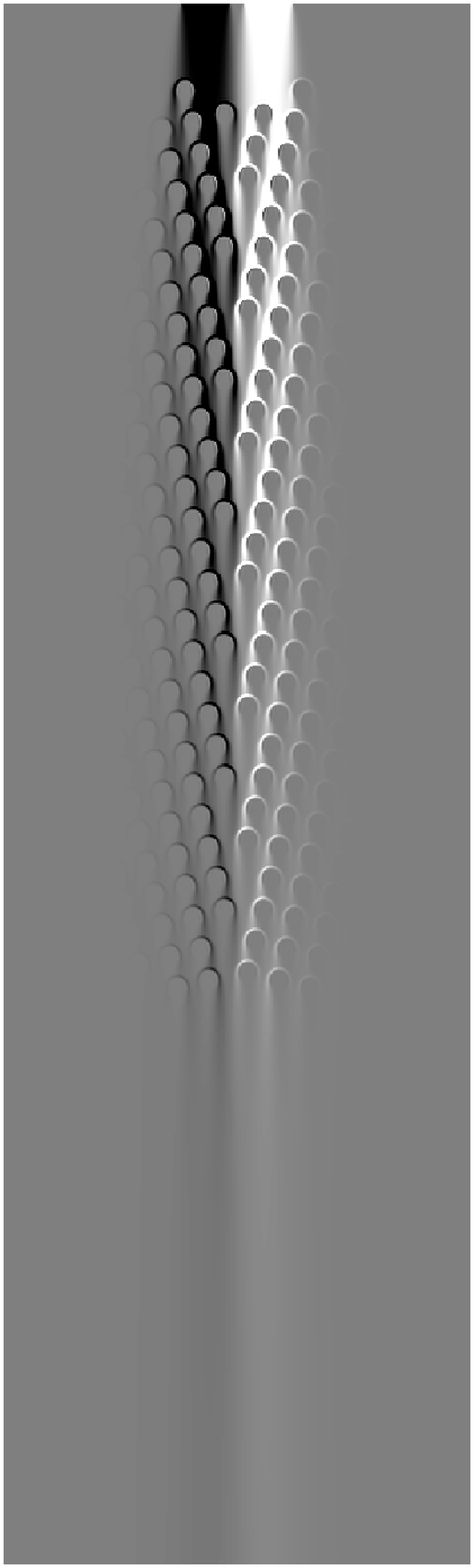}\\
\includegraphics[width = 5cm,angle=90]{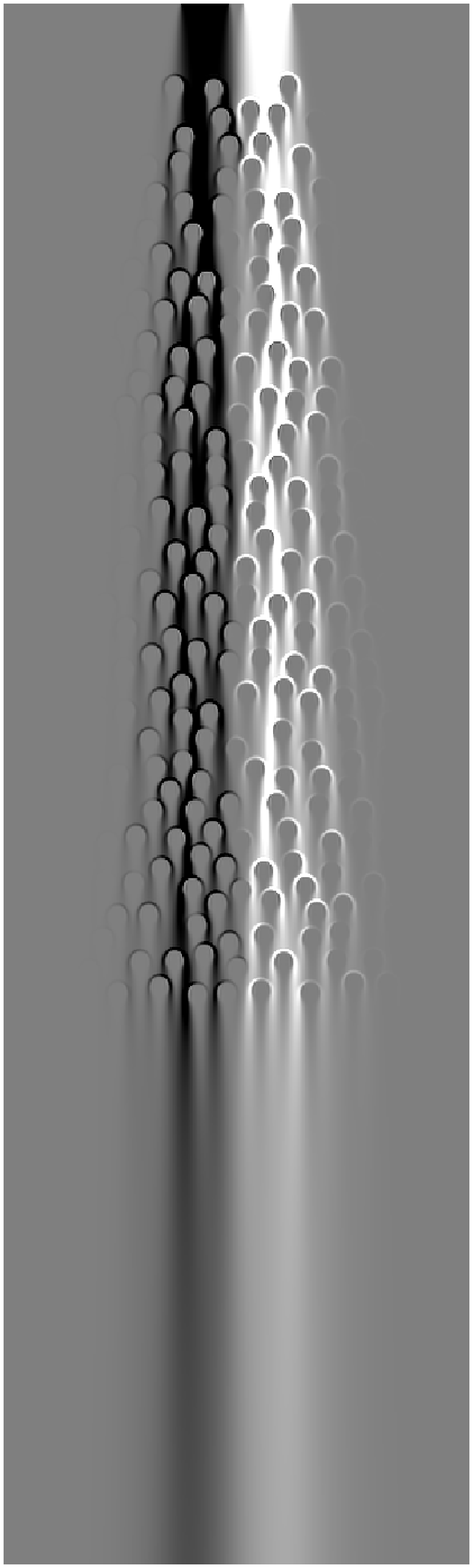}
\end{center}
\caption {Distribution of the two species, black and white,  at the final time $\tau=60$, associated with the periodic pattern (PP) of Fig. \ref{1evol_I} (top) and with a random pattern (RP) of obstacles (bottom). In both cases  ${\hat k}=0.20$ .}
\label{2evol_I}
\end{figure}

We are interested in a number of observables along the flow direction, and not in a direction perpendicular
 to the flow, which we integrate over. The observables we focus on are the local reaction rate $R(x)$,
\begin{equation}
R(x) = k \int dy \, c_a(x, y) c_b(x, y),
\label{Rx}
\end{equation}
and 
 the total  flux as a function of the position $x$, 
\begin{equation}
J(x,t) = \int dy \left[ J_{x,a}(x,y, t) +  J_{x,b}(x, y;t) \right].
\end{equation}
From this quantity  the reaction efficiency at a long time (here taken as $\tau=60$) is evaluated  comparing the fluxes of non reacted particles  at two points, $x={0.5}$ and $x={40}$,
\begin{equation}
\eta(k) = 1 - \frac{J(x=40)}{J(x=0.5)}.
\label{flux}
\end{equation}

In Fig.~\ref{1evol_I} we present the two-dimensional landscape in which the advection, diffusion, and reaction of the two components $A$ and $B$ takes place.
One can see the periodic and tilted structure of the circular obstacles. 
Figure~\ref{2evol_I} (top)  shows the distribution  of both concentrations at time $\tau=60$. 
It is clear that the obstacles focus the concentrations toward the center line, where most of the reaction process takes place. 
This situation should be compared with the case of randomly distributed obstacles,  Fig.~\ref{2evol_I} (bottom).

\section{Analysis of numerical results}
\label{sec3}

\begin{figure}
\begin{center}
\includegraphics[width = 7cm]{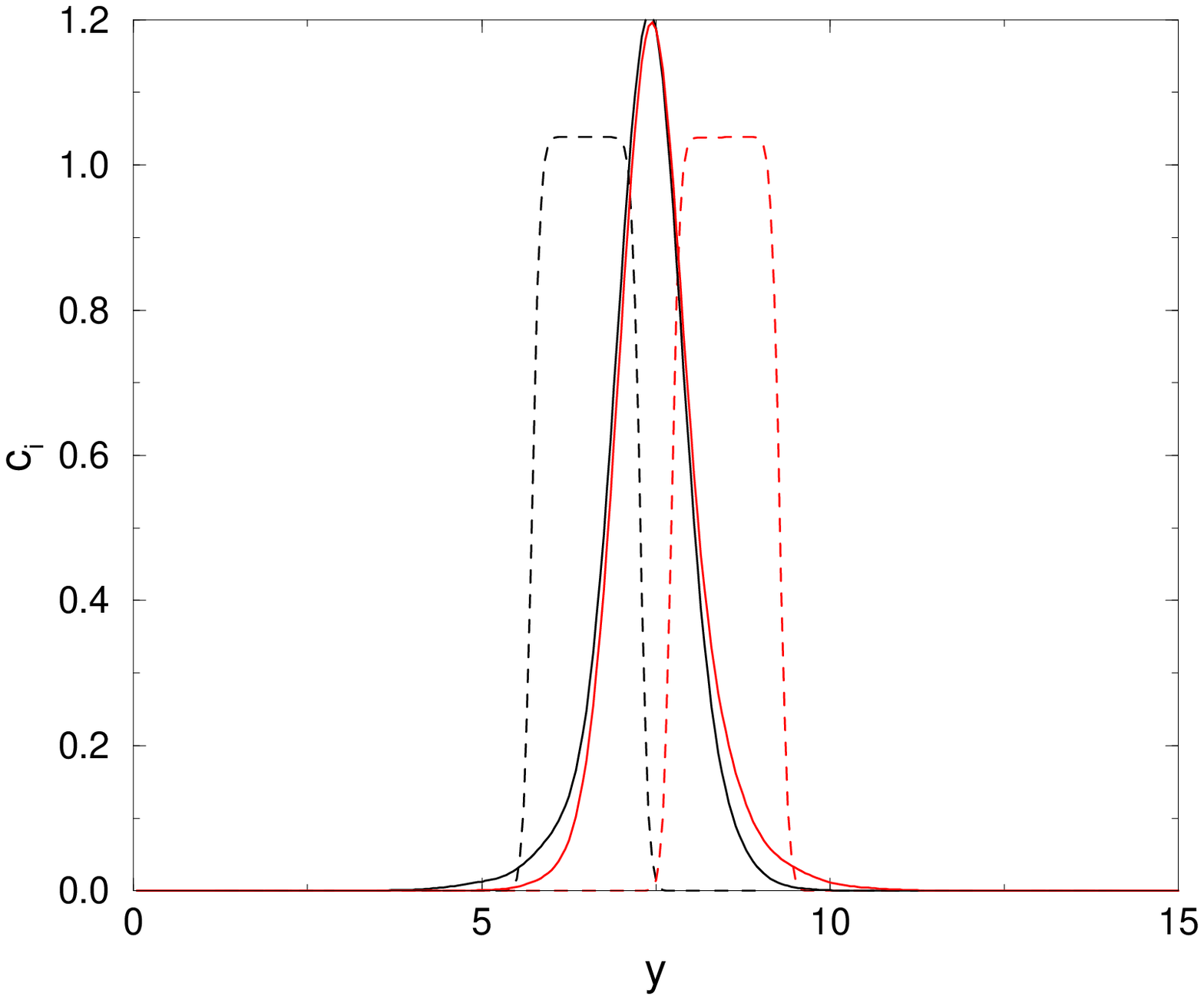} 
\includegraphics[width = 7cm]{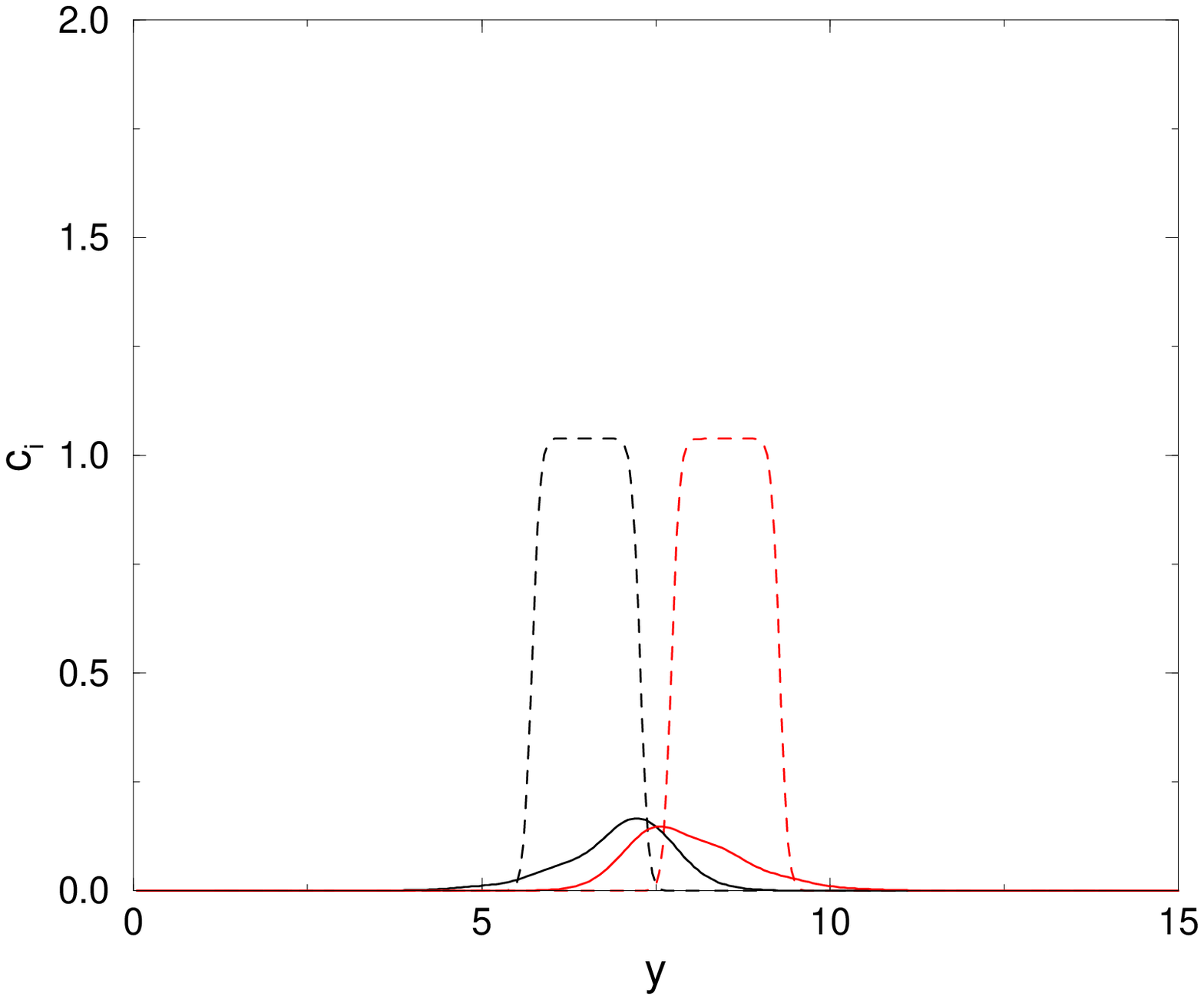} \\
\includegraphics[width = 7cm]{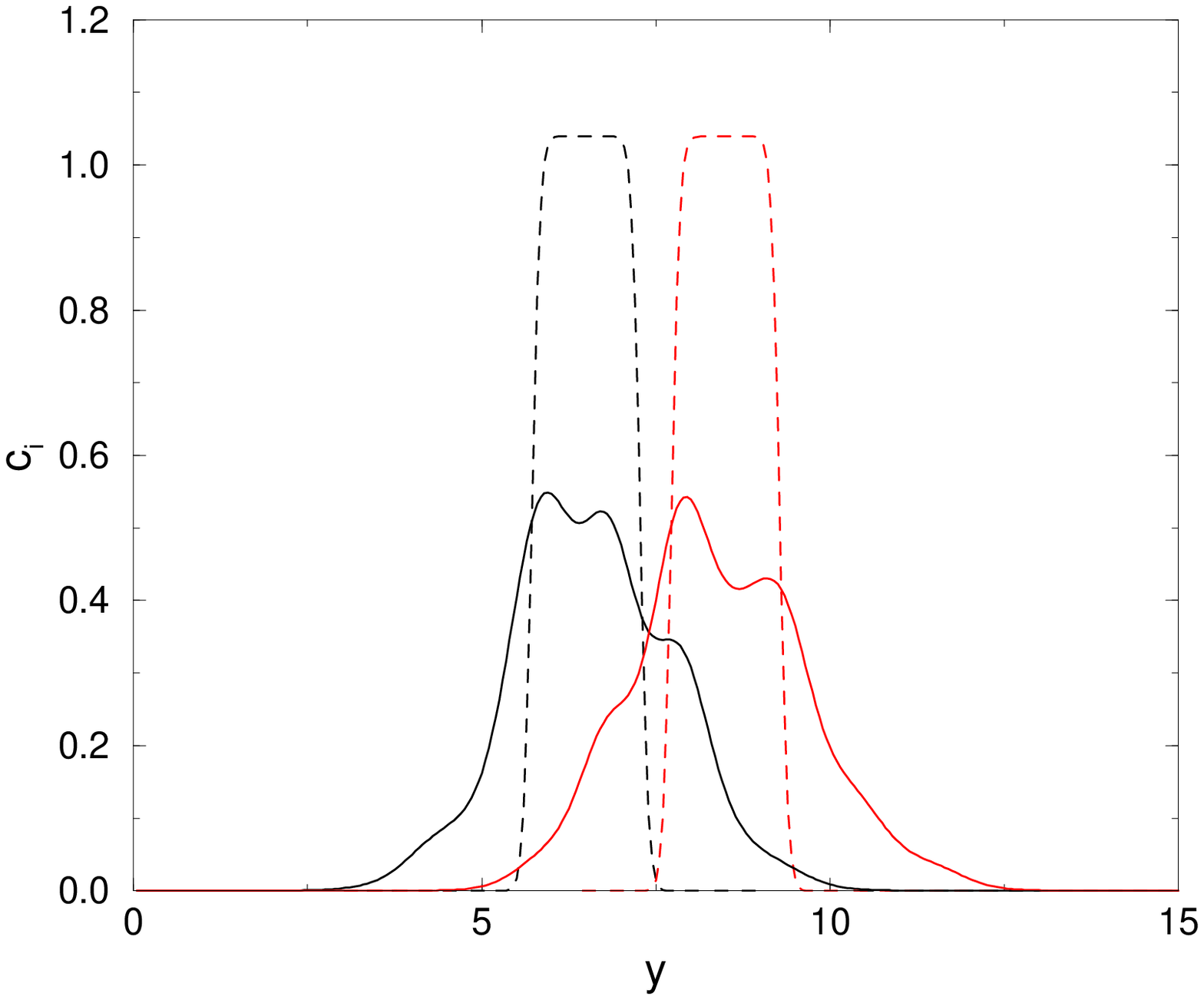} 
\includegraphics[width = 7cm]{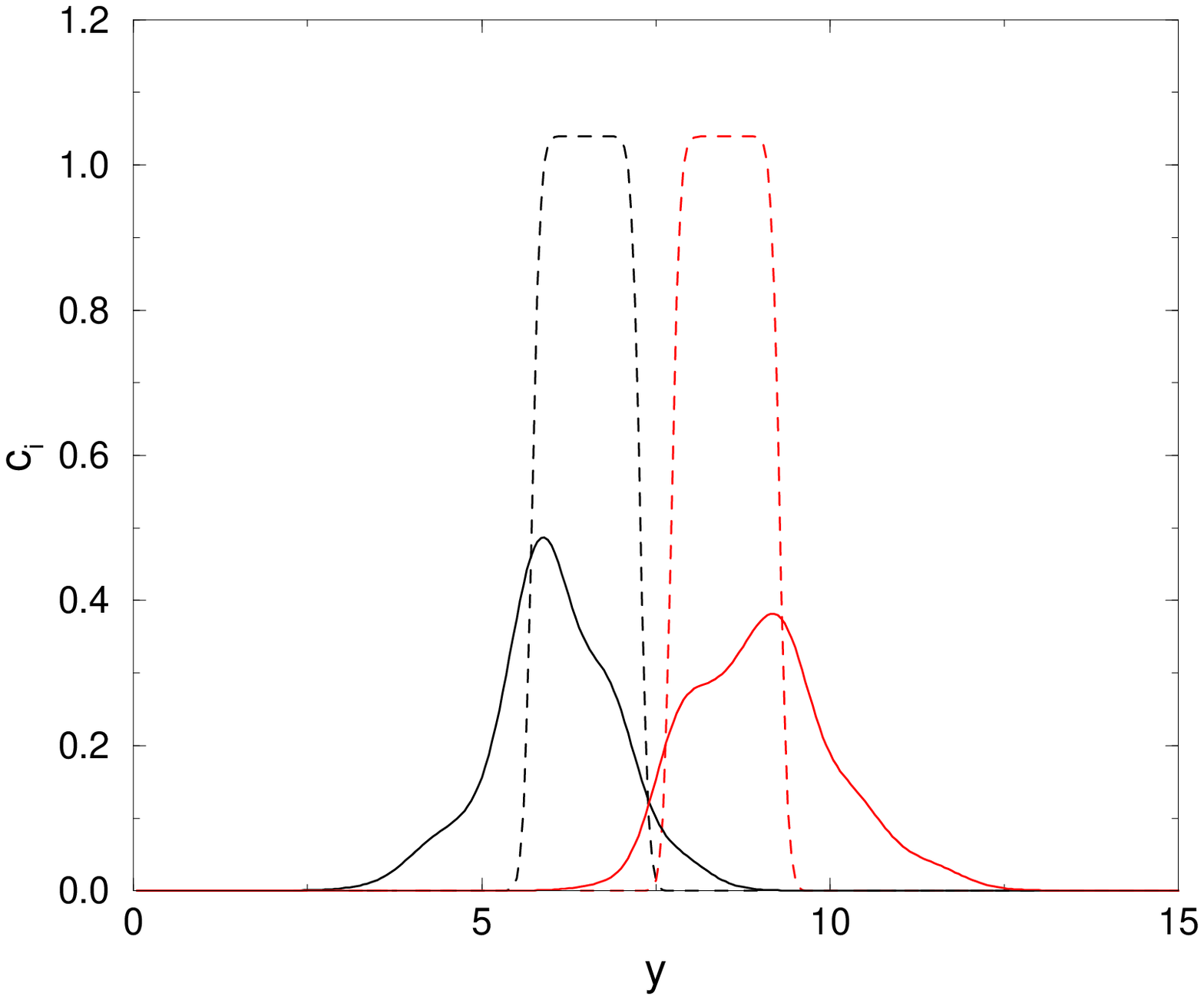}\\
\includegraphics[width = 7cm]{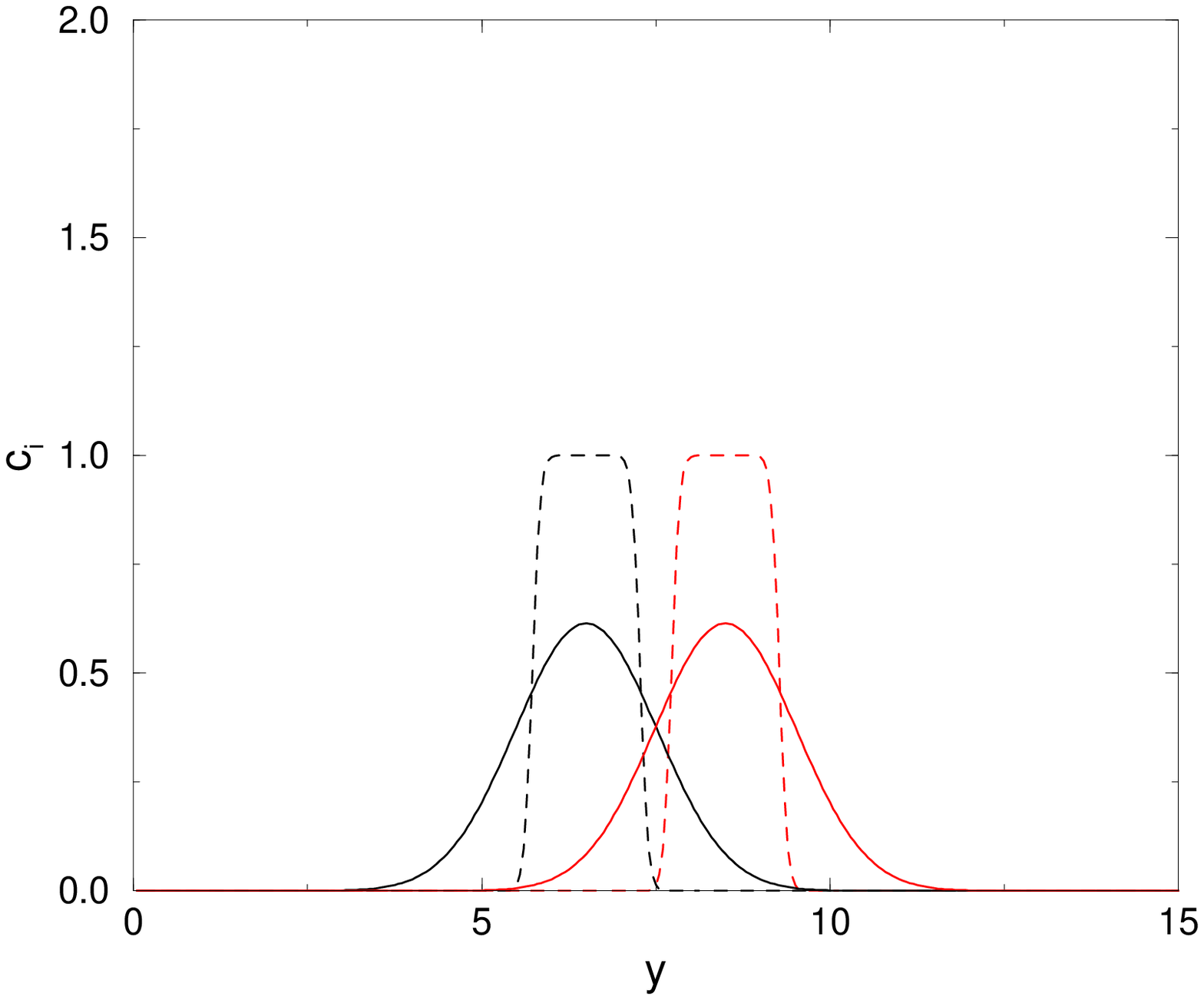} 
\includegraphics[width = 7cm]{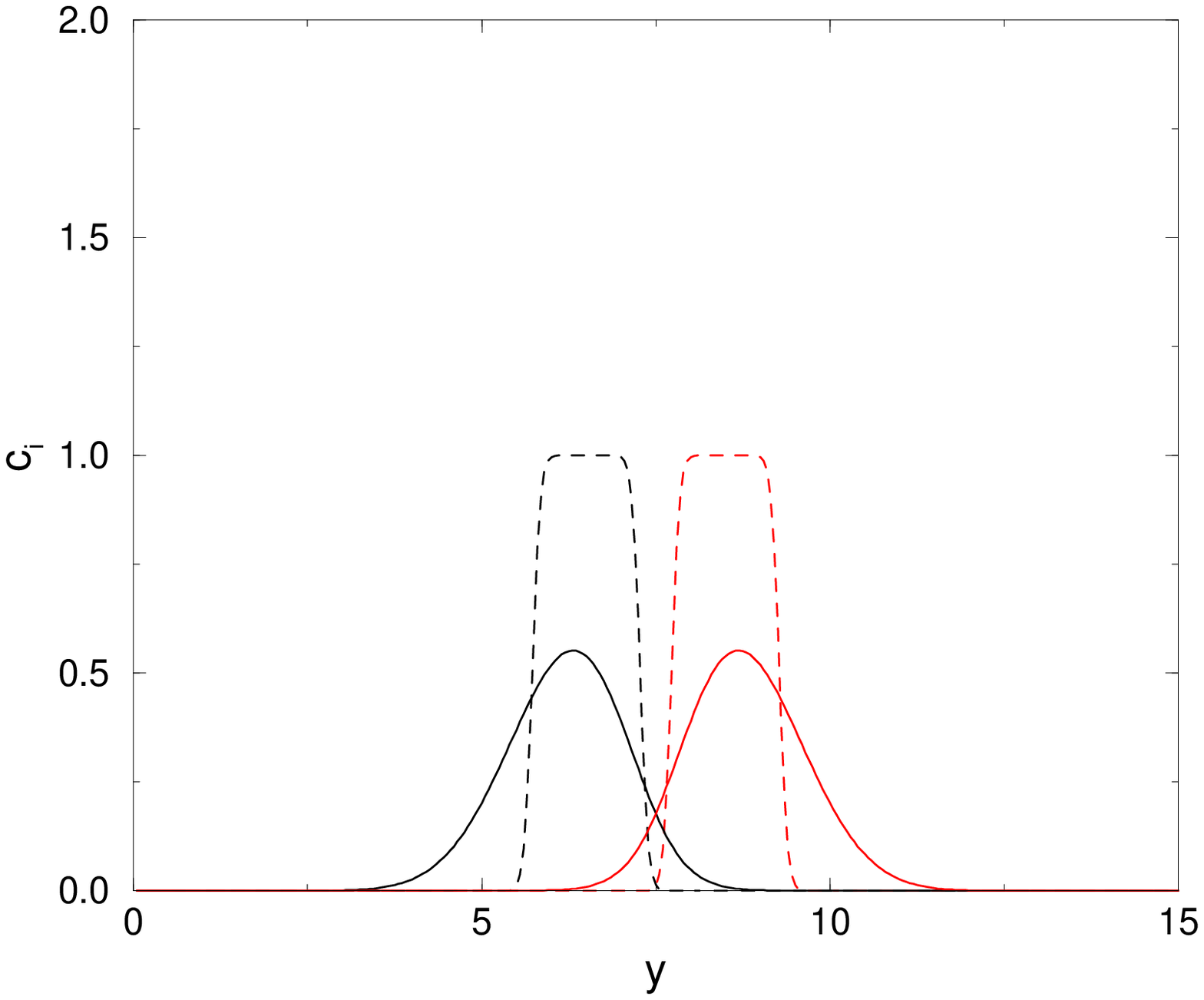}
\end{center}
\caption{Left colum: concentration profiles $c_a$ (black) and $c_b$ (red), at $x=0.5$ (dashed lines) and $x=40$ 
(solid lines) for $\tau=60$, when the species do not react ($\hat k=0$). 
The upper left panel shows the profiles for the periodic pattern (PP), the middle left panel for the random pattern (RP), and the lower left panel when there are no obstacles (NP) at all on the surface. 
Right column: the same three cases (PP, RP, and NP)) as in the left column but now incorporating the reaction with rate coefficient ${\hat k}=0.2$.
}
\label{prof_k0}
\end{figure}

\begin{figure}
\begin{center}
\includegraphics[width = 8cm]{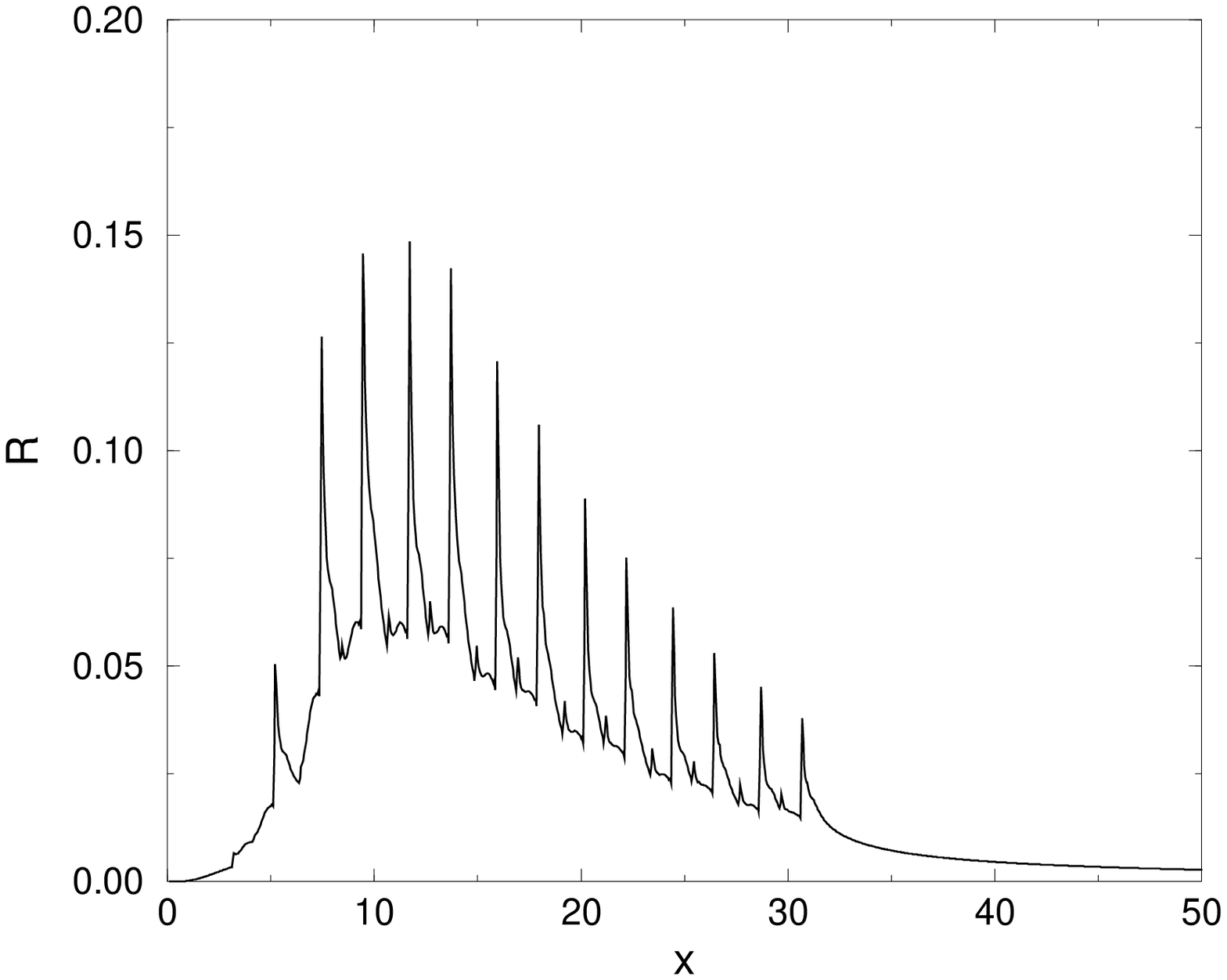}
\includegraphics[width = 8cm]{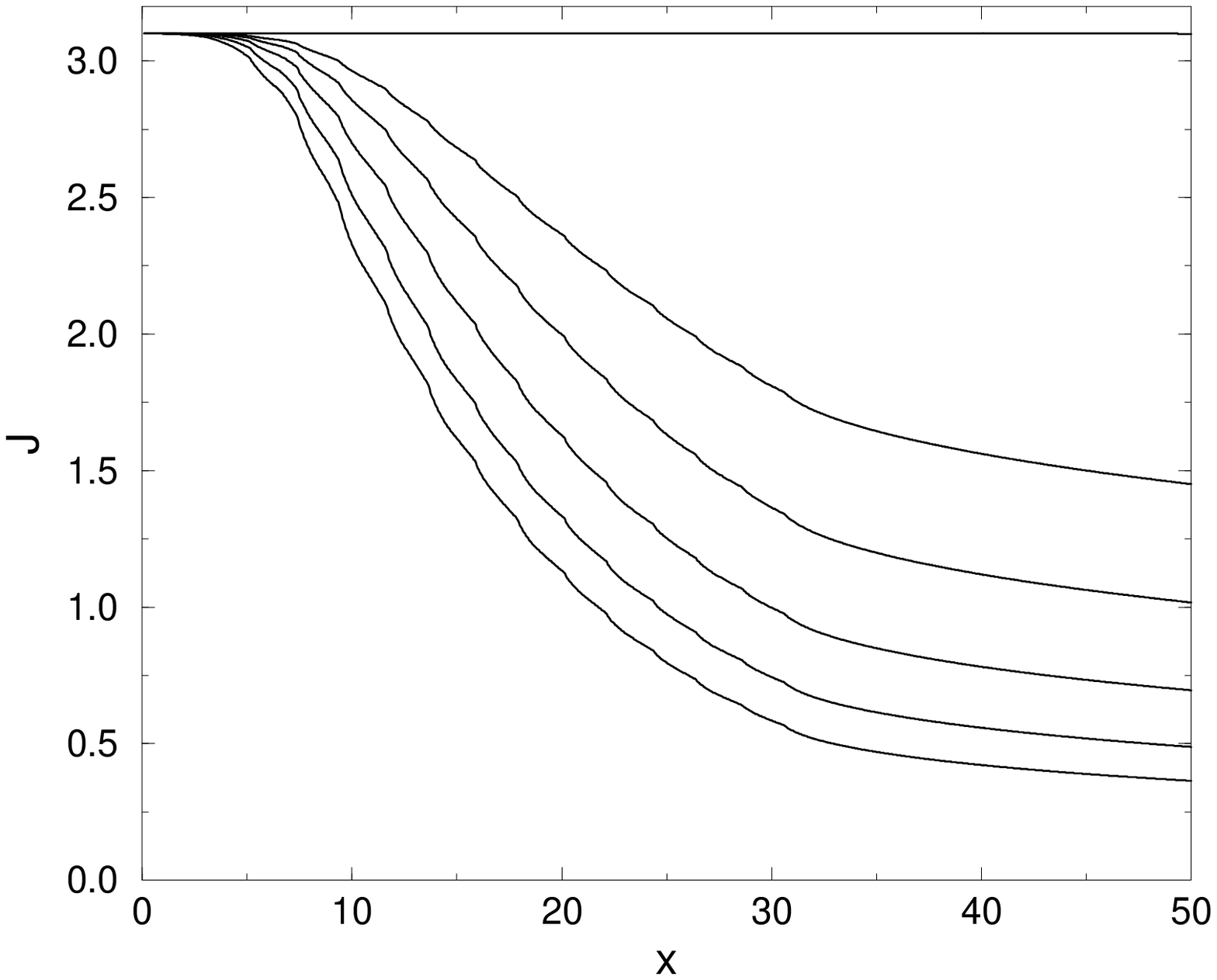} 
\end{center}
\caption{Left panel: reaction rate parameter $R(x)$ [Eq. (\ref{Rx})] for the configuration PP in the steady state 
($\tau=60$) for  ${\hat k=0.2}$.  
Right panel: flux  [Eq. (\ref{flux})]  
for the same case as left panel, for several values of ${\hat k}$. From top to bottom: $\hat k=0, 0.025, 0.05, 0.1, 0.2, 0.4$.
} 
\label{R}
\end{figure}

\begin{figure}
\begin{center}
\includegraphics[width = 10cm]{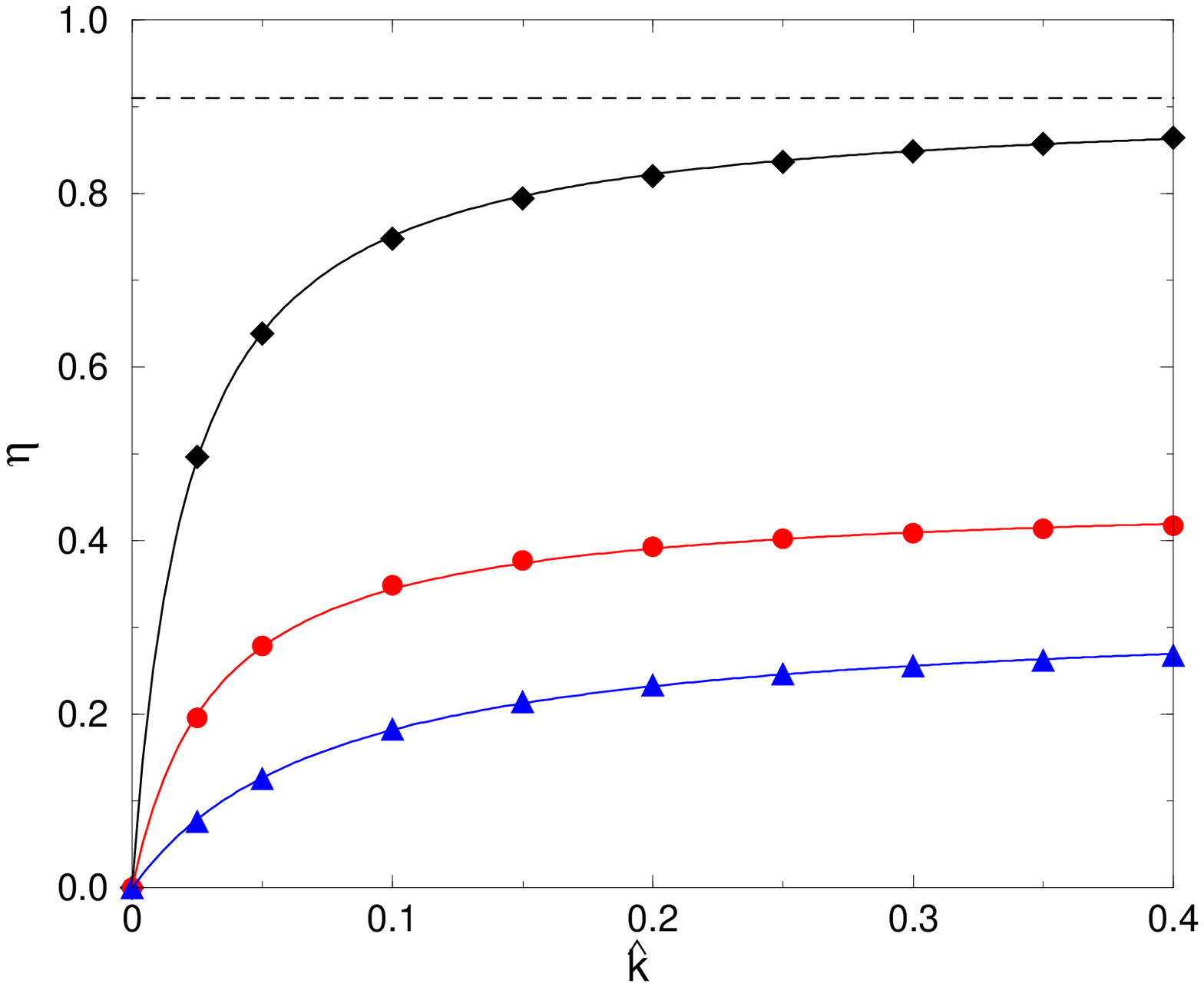}\\
\end{center}
\caption{Reaction efficiency versus reaction rate constant ${\hat k}$  at the final position $x_F=40$ ($\tau = 60$)
for the periodic pattern (PP, diamonds), random pattern  (RP, circles), and without obstacles (NP, triangles).
The dashed line corresponds to the reaction efficiency for perfect mixing, cf. Eq.~(\ref{pmixing}). The full lines are the fittings of Eq.~(\ref{fit}).
}
\label{efficiency}
\end{figure}

In Fig. \ref{prof_k0} we present initial and final  concentration profiles. The left column shows the effect of obstacles on the flow of reactants in the absence of a reaction (${\hat k}=0$). From top to bottom we see the reactant flow patterns when the obstacles are placed in the tilted periodic pattern (PP), in the random placement (RP), and without obstacles (NP). We clearly see the strong focusing effect of the PP geometry compared to the other cases.
In the right column we see the same three cases but now the two species are allowed to react with rate coefficient ${\hat k}=0.2$.  The reaction clearly proceeds much more rapidly when the reactants are focused by the obstacles.  The reaction in the presence of random obstacles and of no obstacles occurs more slowly and at comparable speeds in the two cases, the former slightly more rapidly than the latter.
In Fig. \ref{R} the local reaction rate $R(x)$ given in Eq.~(\ref{Rx}) and the flux $J(x)$ of Eq.~(\ref{flux}) in the steady state are plotted for the focusing geometry (PP) along the flow direction. 
We observe repeated positions at which the reaction is more efficient. These points correspond to rows where the innermost focusing obstacles are closest together. This result opens the possibility of alternative patterns with more active reaction domains.  
The flux is shown for the focusing obstacle geometry for a number of rate coefficients and is seen to be a monotonically decaying function of position. For the focusing obstacle geometry with ${\hat k} =0.2$, for instance, we calculate a reaction efficiency of $\sim 85 \%$, a fairly high value for such a relatively small system. 
In Fig. \ref{efficiency} we explore the dependence of the efficiency on the reaction parameter ${\hat k}$. We see that the efficiency comes close to a maximum value around  ${\hat k}=0.2$, with little improvement above that. This almost-independence of the rate coefficient is an interesting unanticipated feature.
 
It is informative to compare this efficiency with that obtained in the most effective case, that is, that of perfect (totally homogeneous) mixing. In this case, assuming equal initial concentrations $c_a=c_b=c$.  the reaction equation is
\begin{equation}
\frac{ d c}{d \tau} = -{\hat k} c^2.
\end{equation}
Its solution is
\begin{equation}
\frac{c(\tau)}{c(0)} = \frac{1}{1 + c(0) {\hat k} \tau }.
\label{pmixing}
\end{equation}

We choose the comparison time $\tau=40$. If we further choose  $c(0)=1$ and ${\hat k}=0.25$, we find that
$c(40)/c(0) = 1/11$ which gives  an efficiency $\eta =0.91$. This an upper bound for this setup. The focusing configuration thus leads to almost perfect mixing.

We can use this result to obtain an analytic estimate for the efficiency. If we take the velocity of the flowing components 
$\sim 1$ we can substitute time for space,  $\tau=x/v$. We  expect the flux to have a functional form similar to that of the concentration in Eq. (\ref{pmixing}), and thus we propose the following expresion for the efficiency at $x_F$: 
\begin{equation}
\eta(x_F,{\hat k})= \frac{a {\hat k} x_F}{1+ b {\hat k} x_F },
\label{fit}
\end{equation}
with parameters $a$ and $b$ to be fitted.  A nonlinear fit to our numerical data yields  $a=1.085$ and $b=1.195$ (PP),
$a=0.358$ and $b=0.788$ (RP) and $a=0.105$ and $b=0.325$ (NP). In Fig. \ref{efficiency}  we have plotted  
this function for $x_F=40$. This numerical result implies that most of the reacted matter was in the perfect mixing regime characterized by the decay $\sim \tau^{-1}$. This is an unexpected result because of the inhomogeneous concentrations in this system.  It tells us that the reaction is dominated by the small domains near the center of the array where there is quasi-perfect mixing, as is seen in Fig. \ref{2evol_I}.  

\begin{figure}
\begin{center}
\includegraphics[width = 8cm]{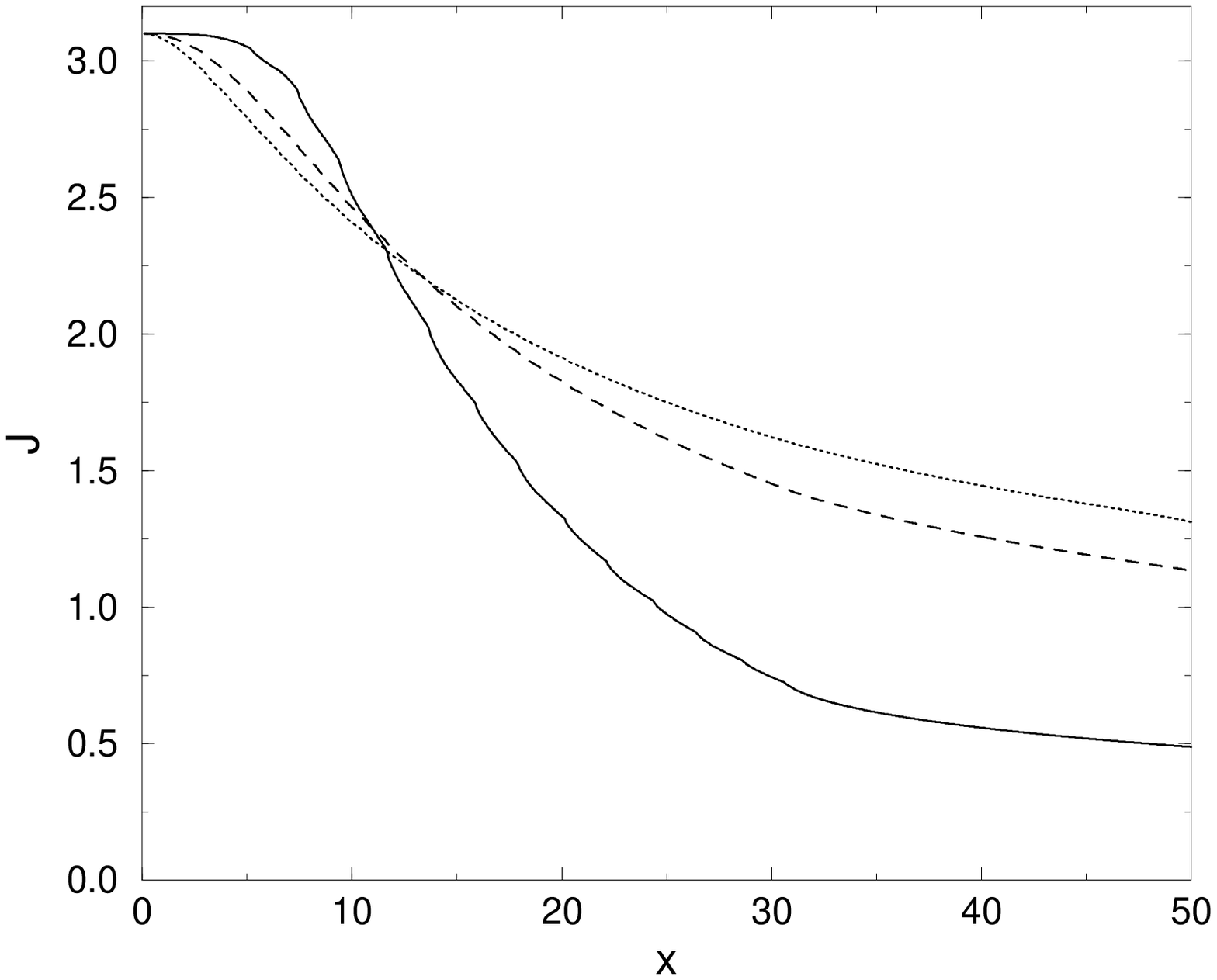}
\includegraphics[width = 8cm]{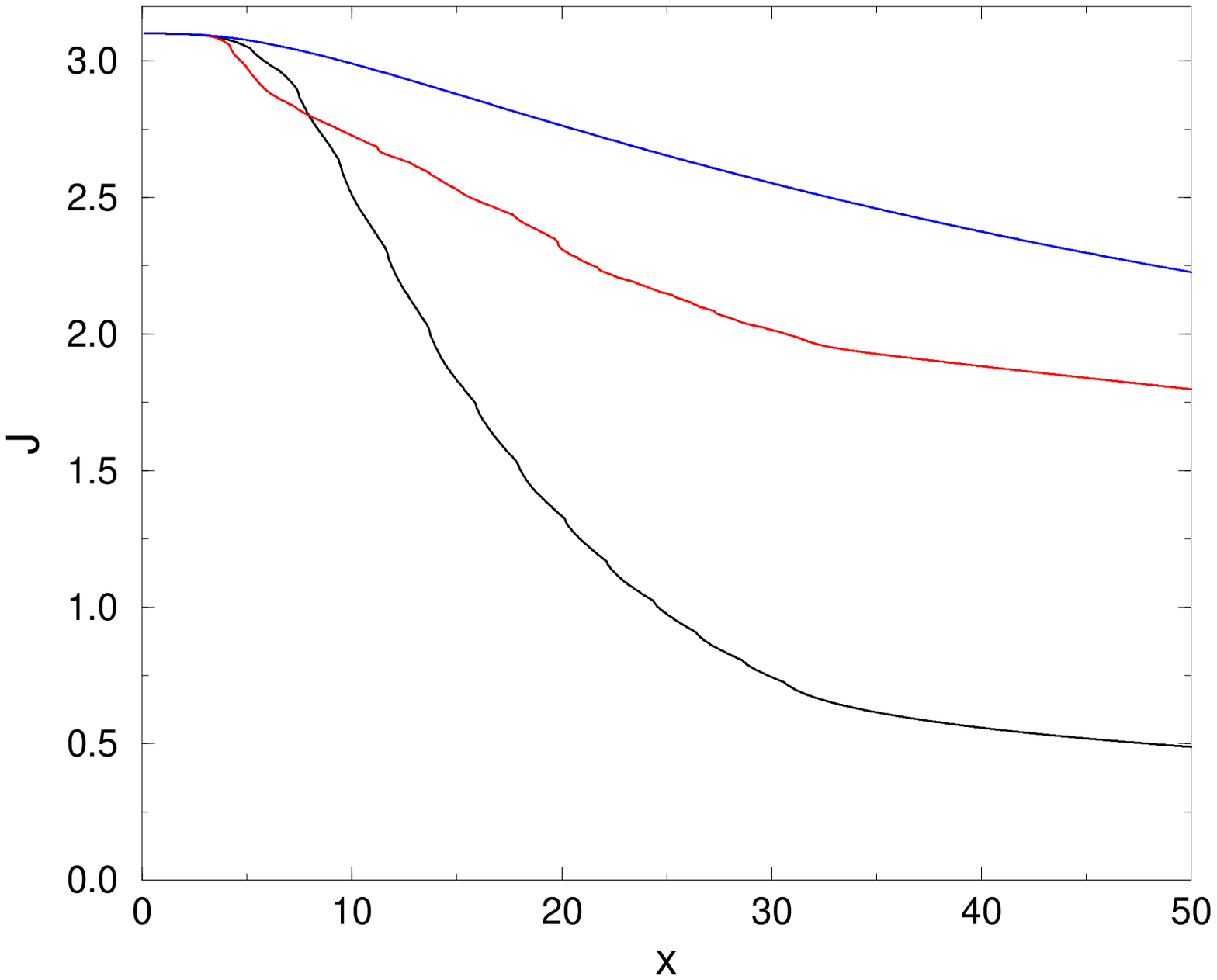} 
\end{center}
\caption{Left panel: plot of the flux for the PP configuration and three values of the diffusion coefficient:
$\hat D=0.01$ (solid line), $0.05$ (dashed line) and $0.1$ (dotted line). 
Right panel: fluxes for the three configurations considered earlier, from top to bottom: NP, RP and PP (same colors as in Fig. \ref{efficiency}).
In both panels $\hat k=0.2$ and $\tau = 60$.
}
\label{diff}
\end{figure}

To complete our analysis, we will explore the effects on these results of varying the diffusion coefficient (assumed to be the same for both species), the focusing pattern geometry, and the sizes of obstacles and particles. 
In the left panel of Fig. \ref{diff} we present the effect of the diffusion parameter. We indicate the focusing domain for the case of our focusing obstacle pattern by dashed lines, that is,  the focusing takes place within this domain. We see that inside this region the reaction is dominated by focusing, but outside of this region the reaction is dominated by diffusion. 
In the right panel of Fig. \ref{diff} we present three fluxes that correspond to different geometries. We see that the flux with the focusing pattern decays faster than the random pattern which is also more effective than no pattern at all.

\begin{figure}
\begin{center}
\includegraphics[width = 10cm]{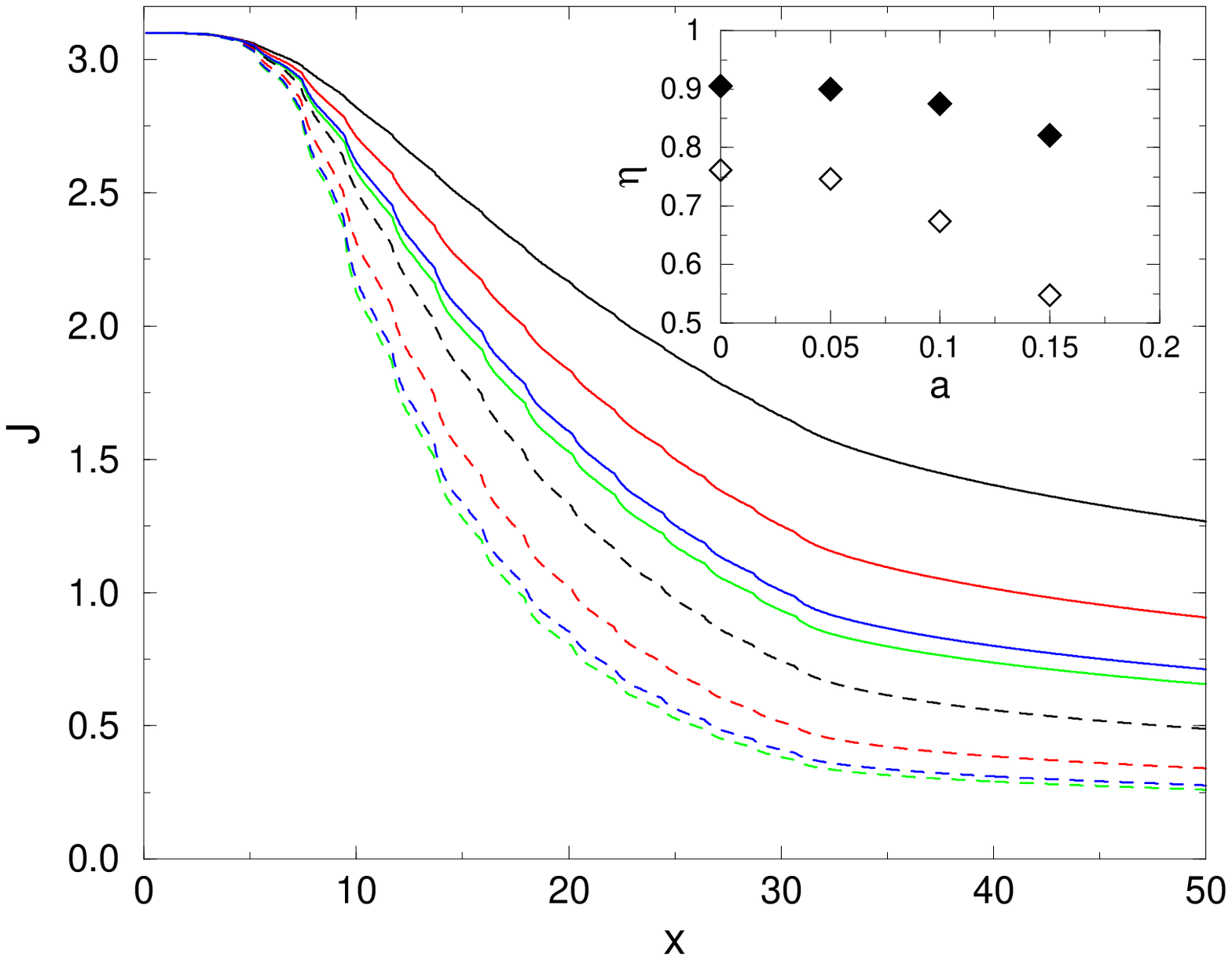}\\
\end{center}
\caption{Comparison of fluxes obtained by using patterns with different  radii of obstacles $a$ and of potential range $d$. 
Solid lines: $d=0.20$; dashed lines: $d=0.25$. For each $d$ value, different values of obstacle radius are represented: 
from top to bottom $a=0.15$ (black), $a=0.10$ (red), $a=0.05$ (blue) and $a=0$ (green, this case corresponds to the $\bf v = \mbox{const.}$ approximation). 
Inset: Efficiency versus parameter $a$ for $d=0.20$ (open symbols) and $d=0.25$ (black symbols).  
}
\label{radius}
\end{figure}
Finally we discuss the effect of the parameters  $d$ (radius of potential interaction) and $a$ (radius of the obstacles) on the reaction efficiency.
From Fig.~\ref{radius} we can see that the stronger effect comes from the change in the parameter $d$. A comparison of pairs of curves with a common value of $a$ shows that the reactant flux  is strongly reduced by increasing  $d$  from 0.20 to 0.25. This reduction in the flux implies an increment in the efficiency (see inset of Fig.~\ref{radius}), as follows.  Returning to Fig.~\ref{lines}, we see that larger $d$ means that more  stream lines of the liquid flow are  affected by the potential, and as a consequence more particles traveling along the stream lines are deviated. Moreover, the deviation of each particle will be greater, that is, particles are moved to further stream lines. An increase of $d$ then clearly produces a larger focussing effect of the reacting particles, and as a consequence the reaction is enhanced. Note that such a change in the parameter $d$ with fixed $a$ corresponds to a change in the particle radius $d-a$ (a change in the reactants) while maintaining the same pattern of obstacles.

Next we fix the parameter $d$ and explore the role of the parameter $a$. In Fig.~\ref{radius} we see that an increase in $a$ reduces the efficiency (see inset). The explanation again lies in the fact that the obstacles lead to a deformation of the stream lines (see Fig.~\ref{lines}), and this effect is stronger for larger $a$. With a stronger deformation the stream lines are displaced further from the obstacles, and as a result more particles can escape from the action of the interaction potential, whose range $d$ is now held fixed. As a consequence the increase of $a$ at fixed $d$ reduces the focussing effect of the pattern and hence the reaction efficiency decreases.

Perhaps most interesting from an experimental point of view is to analyze the effect of varying both $a$ and $d$ while maintaining their difference constant. This corresponds to changing the dimensions of the obstacles, easily modified in an experiment, while keeping the particle radius $d-a$ constant. This would be the scenario in which one attempts to optimize the reaction process for given reactive species. In Fig.~\ref{radius} we can compare the case $d=0.20, a=0.10$ with the case $d=0.25, a=0.15$, which corresponds to changing the obstacle size with a common particle radius of $d-a=0.10$. Results in the figure show that the increase of obstacle size in this example clearly reduces the reactant flux and hence improves the reaction efficiency. 

As a last test, we have compared these results to those of approximation in which the deformation of the fluid flow is neglected, {\it i.e.}, in which the flow velocity is taken to be constant and only the scale $d$ of the interaction potential is taken into account. In Fig.~\ref{lines} this situation would correspond to straight, horizontal stream lines. This approximation was used in our previous work on particle sorting \cite{katja2,ana,James}, and in our present scheme corresponds to the limit $a=0$. In this situation the stream lines are not deformed and hence a larger number of particles are deviated by the potential. In Fig.~\ref{radius} we see the expected result, namely, that this approximation overestimates the reaction efficiency when compared to a finite $a$ value (for a given value of $d$). For the  smallest values of $a$ (obstacles smaller than the particles) this approximation does not qualitatively change the focussing scenario, but as $a$ increases the importance of including the effects of advection clearly increases.

\section{Conclusions}
\label{sec4}

Obstacles placed in carefully selected geometrical patterns have been used in the past to effectively separate colloidal mixtures that are caused to flow over a surface containing such obstacles.  In this paper we have explored the converse, namely, the possibility of speeding up the mixing of components that undergo advective diffusion over a surface containing carefully situated obstacles.  We have illustrated the effects of this mixing by considering the reaction of two species and comparing the reaction rates when the species are allowed to mix by ordinary advective diffusion in the absence and presence of these obstacles.
We have shown that a periodic pattern of tilted obstacles, as opposed to a random placement, is able to effectively focus the streams of reactive species.
 We have furthermore shown that this focusing mechanism leads rather rapidly to reaction rates comparable to those obtained with perfect mixing. We have studied the dependence of the reaction efficiency on the different parameters of the problem. We interpret the focusing mechanism as a consequence of the finite size of the reacting particles. These results could be useful for the design and interpretation of new experiments.

\vskip5mm

This work is supported by Ministerio de Econom\'ia y Competitividad (Spain) through project FIS2012-37655, by the Generalitat de Catalunya through project 2009SGR-878, and by the US National Science Foundation through Grant No. PHY-0855471.

\appendix*
\section{Numerical solution of the Hele-Shaw problem with
circular obstacles}

We will consider a fluid contained between two parallel plates, separated by a
gap $d$ (Hele-Shaw cell). In the limit of small $d$ the velocity field $\mathbf u$
can be considered as two-dimensional, and is given by 
\begin{equation}
 {\mathbf u} = -\frac{d^2}{12\mu} \nabla p,
\label{eq-darcy}
\end{equation}
which is identical to the Darcy's law for the flow in a porous medium. $\mu$ is
the viscosity of the fluid and $p$ is the pressure, which satisfies $\nabla^2 p =
0$.  We place $N$ circular
obstacles or disks of radius $a$ centered at positions ${\mathbf R}_k$, 
$k=1\dots N$.
The velocity of the fluid far from the obstacles is ${\mathbf u}_\infty$. 
At the rigid boundaries
the velocity satisfies the condition of zero normal component, but not the
no-slip boundary condition. The stream lines passing the obstacles are identical
to those of a two dimensional inviscid fluid with the same
geometry \cite{batchelor}.  

The velocity field can be written in terms of the velocity potential $\phi$
as
\begin{equation}
  {\mathbf u} = \nabla \phi, \label{eq-grad}
\end{equation}
which obeys the Laplace equation
\begin{equation}
\nabla^2 \phi = 0.  \label{eq-laplace}
\end{equation}

The general solution of the 2D Laplace equation (\ref{eq-laplace}) can be
written in polar coordinates as
\begin{equation}
 \phi(r,\theta) = \sum_{\lambda=1}^\infty 
\left(a_{\lambda 1} \left(\frac{r}{a} \right)^\lambda +
a_{\lambda 2}\left(\frac{r}{a} \right)^{-\lambda} \right)
\left(b_{\lambda 1} e^{\mathrm{i} \lambda \theta} + b_{\lambda 2} e^{-\mathrm{i}
\lambda \theta} \right),
\label{eq-genlaplace}
\end{equation}
where $a_{\lambda i}, b_{\lambda i}$ are constants that depend on the boundary conditions. 
It is easy to establish that for real $\phi$ these constants should satisfy
\begin{eqnarray}
 a_{\lambda 1} b_{\lambda 1} = (a_{\lambda 1} b_{\lambda 2})^*
\label{eq-condition1}\\
 a_{\lambda 2} b_{\lambda 2} = (a_{\lambda 2} b_{\lambda
1})^*.\label{eq-condition2}
\end{eqnarray}

We will use the multipole expansion (\ref{eq-genlaplace}) for the velocity field
in the
neighbhood of each disk, with the polar coordinates centered at the disk.
Hence we will have $N$ such expansions. The advantage of these expansions
is that the boundary conditions at the disks are easy to formulate. 
In particular, the normal velocity vanishes, that is,
\begin{equation}
\left. \frac{\partial \phi}{\partial r}\right|_{r=a} = 0,
\end{equation}
and then, taking into account the conditions
Eqs.~(\ref{eq-condition1}) and (\ref{eq-condition2}) we get
\begin{eqnarray}
 a_{\lambda 1} = a_{\lambda 2} = 1 \\
 b_{\lambda 1} = b_{\lambda 2}^* \equiv b_\lambda.
\end{eqnarray}
We still have to find one set of $b_\lambda$ constants for each disk $k$
($k=1\dots N$), which we will  denote as $b_\lambda (k)$.

We next consider the ensemble of $N$ disks. All multipole expansions should
simultaneously satisfy boundary conditions on all the discs and at infinity.
It is convenient to employ a complex variable for the position, $\it i.e.,$
we
define the complex position $\tilde r$ as
\begin{equation}
 \tilde r = x  + y\mathrm{i}.
\end{equation}
Henceforth a tilde on any vector will denote a similar definition
as a complex variable. By employing this notation the general solution
Eq.~(\ref{eq-genlaplace}) can be expressed more compactly as a series of powers
of $\tilde r$.

Note that each expansion (\ref{eq-genlaplace}) is valid near the corresponding
disk, but not far from it. In particular the expansions diverge at infinity. The way to
manage an expression for the potential valid at arbitrary distances is to use
only the decreasing powers of the multipole expansions, adding the terms
corresponding to all the disks. Taking into account the boundary condition at
infinity we can then write
\begin{equation}
 \phi({\mathbf r}) = {\mathbf r}\cdot {\mathbf u}_\infty + \sum_{k=1}^N \phi_k ({\mathbf r}),
\label{eq-expN}
\end{equation}
where $\phi_k ({\mathbf r})$ is the distortion of the
potential produced by the
disc $k$, which will have the form
\begin{equation}
 \phi_k({\mathbf r}) = \sum_{\lambda=1}^\infty \left[c_\lambda (k) \left(
\frac{a}{\tilde r_k} \right)^\lambda + c.c. \right].
\label{eq-expNi}
\end{equation}
Here $\tilde r_k = \tilde r - \tilde R_k$ is the position relative to the center of disk $k$, and the $c_\lambda (k)$ are constants.
With this expression the boundary
condition at infinity has already been taken into account. To find the
unknowns $c_\lambda(k)$ we should to make this expansion compatible
with those of each
disk, Eq.~(\ref{eq-genlaplace}), which incorporating boundary contitions can
be written as
\begin{equation}
 \phi({\mathbf r}) =  \sum_{\lambda=0}^\infty 
\left(b_{\lambda} (j) \left(\frac{a}{\tilde r_j} \right)^\lambda +
b_{\lambda}^*(j)\left(\frac{\tilde r_j}{a} \right)^{\lambda} + c.c. \right),
\,\,\, j= 1 \dots N.
\label{eq-exp1}
\end{equation}

To relate the expansion Eq.~(\ref{eq-expN}) with Eq.~(\ref{eq-expNi}) to those of
Eq.~(\ref{eq-exp1}),
it is convenient to use the Taylor expansion of the negative powers of $\tilde r_k$ (position relative to disk $k$) in terms of the positive powers of $\tilde r_j$ (position relative to any other disk $j\neq k$). The
idea is that the positive powers in the local expansion Eq.~(\ref{eq-exp1}) of
any disk should correspond to the terms coming from the rest of the disks in
Eq.~(\ref{eq-expN}). That is, we write
\begin{equation}
 \left(\frac{a}{\tilde r_k}\right)^{\lambda'}=
\sum_{\lambda=0}^{\infty}q_{\lambda`\,\lambda} 
\left(\frac{a}{\tilde R_{k j}}\right)^{\lambda+\lambda'}
\left(\frac{\tilde r_j}{a}\right)^{\lambda},\,\,\,j\neq k,
\label{eq-expanr}
\end{equation}
where $\tilde R_{k j}=\tilde R_j - \tilde R_k$ 
so that $\tilde r_k = \tilde R_{k j} + \tilde r_j$, and the constants
$q_{\lambda`\,\lambda}$ can be obtained from the recurrence relation
\begin{eqnarray}
 q_{\lambda'\,\lambda}&=&
\frac{1-\lambda-\lambda'}{\lambda}q_{\lambda'\,\lambda-1},\nonumber\\
q_{\lambda'\,0}&=&1.
\end{eqnarray}
We now substitute the expansion Eq.~(\ref{eq-expanr}) into Eq.~(\ref{eq-expNi}), to obtain
\begin{eqnarray}
\phi({\mathbf r}) & = & \frac{1}{2} a \tilde u_\infty^*\frac{\tilde r}{a} +
\sum_{\lambda=1}^{\infty}\left[c_\lambda(j)\left(\frac{a}{\tilde
r_j}\right)^\lambda + c.c. \right]  \nonumber \\
& & +\sum_{k\neq j} \sum_{\lambda'=1}^\infty \left[c_{\lambda'}(k)
\sum_{\lambda=0}^\infty q_{\lambda'\, \lambda}
\left(\frac{a}{\tilde R_{kj}}\right)^{\lambda+\lambda'}
\left(\frac{\tilde r_j}{a}\right)^\lambda + c.c. \right],\,\,\,j=1\dots N.
\label{eq-exptot}
\end{eqnarray}

By equating terms of the same power of $\tilde r$ in Eqs.~(\ref{eq-exp1}) and
(\ref{eq-exptot}) we find
\begin{eqnarray}
 b_\lambda(j) & = & c_\lambda(j), \,\,\, \lambda=1\dots N, \label{eq-res00}\\
b_0^*(j) & = & \sum_{k\neq j} \sum_{\lambda'=1}^\infty c_{\lambda'}(k)
\left(\frac{a}{\tilde R_{kj}}\right)^{\lambda'} , \label{eq-res0}\\
b_1^*(j) & = & \frac{a}{2}\tilde u_\infty^* - 
\sum_{k\neq j} \sum_{\lambda'=1}^\infty c_{\lambda'}(k) \lambda'
\left(\frac{a}{\tilde R_{kj}}\right)^{\lambda'+1} , \label{eq-res1}\\
b_\lambda^*(j) & = & \sum_{k\neq j} \sum_{\lambda'=1}^\infty c_{\lambda'}(k)
q_{\lambda'\,\lambda}\left(\frac{a}{\tilde R_{kj}}\right)^{\lambda'+\lambda},
\,\,\,\lambda>1. \label{eq-reslambda}
\end{eqnarray}

These equation can be solved to obtain the constants $c_\lambda(j)$ with which the potential and the velocity field can be calculated through Eq.~(\ref{eq-expN}). In practice one can use a few multipole terms, $\lambda = 1\dots \lambda_{max}$, for a moderate value of $\lambda_{max}$ since the series converge rapidly (more terms are necessary as the distances between disks are decreased). 
The procedure involves  finding $c_\lambda(j)$ only once, and permits us to find fluid velocity at any position by only summing a few contributions from each disk.

A dipolar approximation ($\lambda_{max}=1$) provides reasonably good results if the disks are not very close to each other, and its solution can be used as an initial step of an iterative solution of the complete problem. This solution reads:
\begin{equation}
 c_1(j) \simeq \frac{a}{2} \tilde u_\infty - 
\frac{a}{2} \tilde u_\infty^* \sum_{k\neq j}
\left(\frac{a}{\tilde R_{kj}}\right)^2 \,\,\,\lambda=1\dots N.
\label{eq-dipolar}
\end{equation}
Then the velocity field in this approximation is
\begin{eqnarray}
 u_x({\mathbf r}) \simeq u_{x\, \infty} - 
a \sum_{j=1}^N \left( \frac{c_1(j)}{\tilde r_j^2} + c.c. \right) 
\nonumber \\
 u_y({\mathbf r}) \simeq u_{y\, \infty} - 
a \sum_{j=1}^N \left( \frac{c_1(j)}{\tilde r_j^2}\mathrm{i} + c.c. \right).
\end{eqnarray}

Finally, the system of equations (\ref{eq-res0})-(\ref{eq-reslambda}) takes the following form:
\begin{eqnarray}
 c_1(j) & = & \frac{a}{2}\tilde u_\infty - 
\sum_{k\neq j} \sum_{\lambda'=1}^\infty c_{\lambda'}^*(k) \lambda'
\left(\frac{a}{\tilde R_{kj}^*}\right)^{\lambda'+1} , \label{eq-sist1}\\
c_\lambda(j) & = & \sum_{k\neq j} \sum_{\lambda'=1}^\infty c_{\lambda'}^*(k)
q_{\lambda'\,\lambda}\left(\frac{a}{\tilde R_{kj}^*}\right)^{\lambda'+\lambda},
\,\,\,\lambda>1. \label{eq-sistlambda}
\end{eqnarray}
With these,  the velocity field is given by
\begin{eqnarray}
 u_x({\mathbf r}) = u_{x\,\infty} - \sum_{j=1}^N  \sum_{\lambda'=1}^\infty 
\left( \lambda c_\lambda(j) \frac{a^\lambda}{(\tilde r - \tilde R_j)^{\lambda+1}} + c.c \right) \nonumber \\
 u_y({\mathbf r}) = u_{y\,\infty} - \sum_{j=1}^N  \sum_{\lambda'=1}^\infty 
\left( \lambda c_\lambda(j) \frac{a^\lambda}{(\tilde r - \tilde R_j)^{\lambda+1}}\mathrm{i} + c.c \right).
\end{eqnarray}

To solve Eqs.~(\ref{eq-sist1}) and (\ref{eq-sistlambda}) an iterative method can be used. These equations can be formulated as a matrix relation,
\begin{equation}
 C = A C^* + B,
\end{equation}
where $C$ is a vector containing all the $c_\lambda(k)$ coefficients. Then we can apply the following iteration, which can be shown to be convergent:
\begin{equation}
 C_i = A C_{i-1}^* + B,
\end{equation}
with the initial term $C_0$ given by the dipolar approximation 
Eq.~(\ref{eq-dipolar}).

\end{document}